\documentclass[10pt,aps,prx,showpacs,superscriptaddress,reprint,citeautoscript]{revtex4-1}   

\usepackage{amssymb,amsmath,amsfonts,bm} 
\usepackage{graphicx} 
\usepackage[rgb]{xcolor}  
\usepackage{subcaption}
\usepackage{dcolumn}   
\usepackage{bm}        
\usepackage{url}

\usepackage{hyperref}
\usepackage{pstricks} 

\def\eeq{\relax}
\def\beq#1#2\eeq{\begin{equation}\label{#1}#2\end{equation}}
\def\bal#1#2\eal{\begin{align}\label{#1}#2\end{align}}
\def\bse#1#2\ese{\begin{subequations}\label{#1}#2\end{subequations}}
\def\ba{\begin{aligned}}   \def\ea{\end{aligned}}

\def\XXint#1#2#3{{\setbox0=\hbox{$#1{#2#3}{\int}$}
\vcenter{\hbox{$#2#3$}}\kern-.5\wd0}}

\newcommand{\E}{\ensuremath{\mathrm{e}}}
\def\Re{\operatorname{Re}} 
 
\newcommand{\w} {v}

\def\dd{\operatorname{d}} 
\def\sgn{\operatorname{sgn}}

\def\ZL#1{\textcolor{blue}{#1}}

\begin{document}

\title{Unilateral and nonreciprocal transmission through bilinear spring systems}
\author{Zhaocheng Lu}
\author{Andrew N. Norris}
\affiliation{Mechanical and Aerospace Engineering, Rutgers University, Piscataway, NJ 08854-8058 (USA)}
\date{\today}

\begin{abstract}
Longitudinal wave propagation is considered in a pair of  waveguides connected by  bilinear spring systems.  The nature of the  nonlinearity causes the compressive and tensile force-displacement relations of the bilinear spring to behave in a piece-wise linear manner, and all transmitted and reflected waves scale linearly with the incident wave amplitude. We first concentrate on a single bilinear spring connecting two waveguides.  By controlling the bilinear stiffness parameters it is possible to convert a time harmonic incident wave into a transmitted wave of the same period but with particle displacement of a single sign, positive or negative, an effect we call unilateral transmission.  Nonreciprocal wave phenomena are obtained by introducing spatial asymmetry.  A simple combination of a single bilinear spring with a mass and a linear spring  shows significant nonreciprocity with transmission relatively high  in one direction and low in the opposite direction. 
\end{abstract}

\maketitle

\section{Introduction}\label{sec1}  
Reciprocity is a fundamental physical principle of wave motion in the presence of time-reversal symmetry: the same incident wave traveling in opposite directions should result in the same transmitted wave.  However, reciprocity limits the control over wave propagation.  Violation of this principle can enable a variety of useful and tunable nonreciprocal wave dynamics such as transmission manipulation \cite{cummer2014}, energy localization \cite{Wang2020}, phase shifters \cite{Hamoir2012,Palomba2018} and topological protection \cite{Lu2014}.

The diode-like component which shows one-way acoustic or elastic wave propagation is the most fundamental nonreciprocal application.  Active approaches to breaking reciprocity  operate by either introducing moving flow into the propagation medium \cite{Zangeneh-Nejad2018,Fleury2014}, or applying spatial and/or temporal modulations of the medium properties \cite{Nassar2017a,Nassar2018a,Nassar2017}.  However, the first type of active methods could result in phase shifts \cite{Zangeneh-Nejad2018} or power splitting \cite{Fleury2014}, which are not the properties of diode; The second approach usually modulates the whole propagation medium and then takes advantage of the modifications in  dispersion relations, operating in a different manner from diode-like component.  Similar to their electronic counterpart, the acoustic and elastic diodes usually have a spatially compact region  to achieve the one-way energy transmission.  Passive nonlinear systems provide a practical  alternative.  

Passive violation of reciprocity requires a departure from linearity combined with spatial asymmetry of the system.  
The first significant passive acoustic diode was a nonlinear medium attached to a linear periodic waveguide \cite{Liang2009}, designed so that the nonlinear part generates the second harmonic falling in the bandpass of linear periodic waveguide.  The first bifurcation-based acoustic rectifier and switch were experimentally demonstrated in a granular chain with a point defect to generate waves at lower frequencies in the bandpass of the structure \cite{Boechler2011}.  {However, this rectification mechanism is only clearly evident with a defect placed at certain location, which limits the application of the design for sound and vibration isolation.}  Recent works have overcome this with  systems composed of two linear media connected by a compact nonlinearity.  The spatially asymmetric nonlinear part can be a  cubic spring-mass chain \cite{Darabi2019}, a linear array of  spherical granules 
\cite{Zhang2019c} or vibro-impact induced elements with unequal grounding springs \cite{Grinberg2018}. However, all of these nonlinear systems are highly amplitude-dependent.  

Here we propose an amplitude-independent diode-like structure comprising two semi-infinite bars connected by a spatially asymmetric bilinear spring-mass system.  The bilinear spring displays different linear load-deformation relations depending on whether the deformation is a state of compression or extension.  This simple either-or nonlinearity has the unique and important property that the response scales linearly with  the amplitude of the incident wave.  However, the discontinuous piecewise linear constitutive relation is a strong nonlinearity making it difficult to find analytical solutions as compared with weakly nonlinear models for which perturbative methods can be used.  Examples of the latter include cubic nonlinearity \cite{Narisetti2010} and Hertzian normal contact \cite{Narisetti2012}.  However, as we show here it is possible to find a semi-analytical solution for nonreciprocal wave transmission and reflection.  To realize the geometric asymmetry, we simply use a bilinear and a linear spring in series and connected by a mass.  This idea comes from our previous bilinear spring-mass chain system with spatial stiffness modulation in \cite{Lu2019}, showing nonreciprocal pulse propagation.  
Additionally, we present an interesting reciprocal 
phenomenon of  an oscillatory incident wave converted into a transmitted wave  with particle displacement of a single sign.  

The outline of this paper is as follows.  Section \ref{sec2} concentrates on the case of  a single bilinear spring  mechanically coupling  two semi-infinite waveguides.  The system is nonlinear but still reciprocal.  However, an interesting phenomenon of unilateral transmission is introduced.  The diode-like property is then achieved by introducing spatial asymmetry in the system.  Section \ref{sec4} discusses the nonreciprocal case where the connection is a simple chain of a mass and two springs, one  bilinear and one linear.  Significant nonreciprocal transmission is demonstrated using computational and semi-analytical methods.  Section \ref{sec5} concludes the paper.

\section{Pulse transmission through a bilinear spring}\label{sec2}

\begin{figure}[h]
\centering
\includegraphics[width=0.80\linewidth]{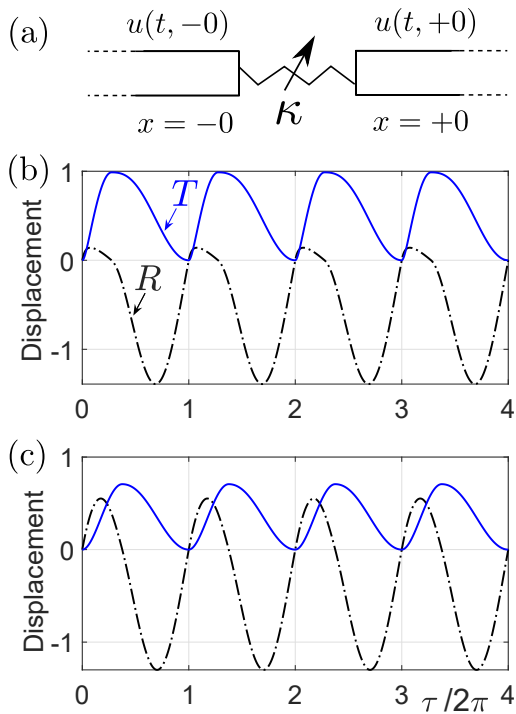}
\caption{Reciprocal model and transmission design.  (a) Model of two semi-infinite waveguides connected by a bilinear spring.  (b)-(c) Examples of transmission coefficients with parameters satisfying equation \eqref{4-}, ensuring that $T\ge 0$.  Note that the maximum value is $T_\text{max} = f(\tau_+)$ since $T+R=f$ and $R=0$ at \ZL{non-dimensional time} $\tau = \tau_+$.  (b) \ZL{$\tau_+ / 2 \pi = 0.275$}, $(\alpha_+ , \alpha_-)=\omega(0.2728,\, 6.3130)$.  (c) $\tau_+ / 2 \pi = 0.375$, $(\alpha_+ , \alpha_-)=\omega(0.2257,\, 0.8349)$. }
\label{fig_1}
\end{figure}

We begin with the case of a single bilinear spring coupling two semi-infinite one-dimensional waveguides (bars), Figure \ref{fig_1}(a).  \ZL{In this paper, we suppose that the spring is much smaller than the wavelength, and hence its size can be ignored}.  The displacements in the bars are  
\beq{u_xpm}
u(t,x) = \begin{cases}
f(t-\frac xc )+R(t+\frac xc ), & x<0 , 
\\
T(t-\frac xc ), & x >0, 
\end{cases}
\eeq
where $f$, $R$ and $T$ represent the incident, reflected and transmitted waves, respectively, and $c$ is the wave speed.  For the moment the incident wave is an arbitrary pulse defined by the \ZL{differentiable } function $f\in C^1$ . 

The bilinear spring force relation is, see Fig.\ \ref{fig_1}(a), 
\beq{F}
F(t,\pm 0) = \kappa [u] \, ,
\eeq
where $[u]$ is the extension, 
\bal{u_fRT}
[u] & \equiv u(t,+0) - u(t,-0) \notag
\\ & = T-R-f \, , 
\eal
$\kappa$ is the bilinear spring stiffness 
\beq{kappa_upm}
\kappa = \begin{cases}
\kappa_- , & [u] < 0 \, , 
\\
\kappa_+ , & [u] > 0 \, .
\end{cases}
\eeq
The stiffness $\kappa_-$ for $[u]<0$ is associated with a compressive (negative) stress at the interface, while $\kappa_+$ for $[u]>0$ corresponds to a tensile (positive) stress.  $F(t,\pm 0)$ is the force in the continuous medium on either end of the spring.  The bars are similar with  Young's modulus $E$ and cross-section $A$, so that 
\beq{F2}
F(t,x)= EA\frac{\partial u}{\partial x}.
\eeq
Equation \eqref{F} implies that the  spring at $x=0$ connecting the two semi-infinite bars  experiences the same force on either end, $F(t,+0) = F(t,-0)$.  Hence, 
$
R' - f' = -T' $, 
which can be integrated to yield
\beq{fRT2} 
R + T = f \, . 
\eeq

Combining Eqs.\ \eqref{F} \eqref{u_fRT} 
and \eqref{fRT2}, using $F(t,+0) = -  T'(t)EA/c $,  implies an ordinary differential equations for $R$, 
\beq{govern_1}
R' + \alpha R = f' \, ,
\eeq where
\beq{alpha}
\alpha  = \frac{2c \kappa}{EA} = \begin{cases}
\alpha_- , & [u]<0 , 
\\
\alpha_+ , & [u] >0. 
\end{cases}
\eeq
Noting from Eqs.\ \eqref{u_fRT} and \eqref{fRT2} that $[u] = -2R$, it follows that the value of  $\kappa$ (and $\alpha$) depends on $\sgn (-R)$ and switches at times when $R$ is zero.  In summary, 
\beq{R}
\left. 
\begin{matrix} R>0
\\  R<0 
\end{matrix}
\right\} \ \Leftrightarrow \ 
\begin{cases} \text{spring is compressed}, \ \alpha = \alpha_- ,
\\  \text{spring is extended}, \ \alpha = \alpha_+.
\end{cases}
\eeq

Equation \eqref{govern_1} may be integrated for $R$ in an interval of time where $\alpha$ is single valued, which is appropriate to the bilinear spring.  The reflection $R$ is therefore a sequence of solutions for the purely linear system, stitched together at the instances when $R$ (and $[u]$) changes sign.  Let $t_j$, $j\in \mathbb N $, be such a time, i.e.\ $R(t_j)=0$, then in the interval $t\in (t_j , t_{j+1})$ for which $R$ subsequently is of one sign, positive or negative, we have 
$T = f-R $ and  
\beq{6} 
R= f-
  \E^{-\alpha  (t-t_j)}f(t_j) - 
 \alpha \E^{-\alpha t}\int_{t_j}^t \E^{\alpha s} f(s) \dd s . 
\eeq
The sign of $R$ in this interval depends upon the derivative of $R$ at $t=t_j$.  Hence from \eqref{govern_1}, $\sgn R' = \sgn f'(t_j)$ in this interval. 
Note that eqs.\ \eqref{fRT2} and \eqref{govern_1} together imply  
\beq{TR}
T' = \alpha R \, , \, \text{and} \ T'' = \alpha f' \ \text{when} \ R=0 \, .
\eeq
It follows that zeroes of $R$ correspond to stationary points of $T$ and the value at the stationary point is $T(t_j) = f(t_j)$.  Furthermore,  $T$ is a local maximum (minimum) at a zero of $R$  if $f'$ is negative (positive). 

We now specialize the incident wave to be time harmonic, starting at $t=0$. Before considering the bilinear model, it is instructive to first examine the linear spring. 

\subsection{Linear spring}
Assuming a sinusoidal incident wave of frequency $\omega$, introduce the non-dimensional time $\tau = \omega t$ so that $f(t) \to f(\tau)$, 
\beq{7}
f(\tau) = H(\tau)
\sin \tau \, ,
\eeq
and $H$  is the Heaviside step function.
The response is 
\beq{29}
\ba
R &= \big[ R_\text{ss} (\tau,\theta) + R_\text{tr}(\tau,0,\theta) \big]H(\tau) ,
\\
T &= \big[ T_\text{ss} (\tau,\theta) + T_\text{tr}(\tau,0,\theta) \big]H(\tau) .
\ea
\eeq
Here "ss" is the steady state solution and "tr" the transient required to satisfy the initial conditions at $\tau =0$, with 
\beq{30}
\ba 
R_\text{ss}(\tau,\theta) &= \cos \theta \sin (\tau + \theta), 
\\
T_\text{ss}(\tau,\theta) &= -\sin \theta \cos (\tau + \theta), 
\ea
\eeq
where \ZL{$\theta$ is parameter used to satisfy relations such as eqs.\ \eqref{fRT2} and \eqref{govern_1}}
\beq{9}
\tan \theta = \frac{\alpha}{\omega} \, ,
\eeq
and 
\beq{31}
\ba 
R_\text{tr}(\tau,\tau_0,\theta) &= - R_\text{ss}(\tau_0,\theta) \, \E^{-(\tau-\tau_0)\tan \theta} \, ,
\\
T_\text{tr}(\tau,\tau_0,\theta) &= - T_\text{ss}(\tau_0,\theta) \, \E^{-(\tau-\tau_0)\tan \theta} \, .
\ea
\eeq
\ZL{Note that $\tau_0 = 0$ in eq.\ \eqref{29} but} we include the dependence on the parameter $\tau_0$ in eq.\ \eqref{31} for later use.  

The exponential decay of the transient implies that the solution \eqref{29} quickly tends to the steady state  $R= R_\text{ss}$, $T= T_\text{ss}$. 
In the steady state limit the scattered energy flux averaged over a cycle equals the averaged incident flux, leading to the  energy conservation relation  
\begin{equation}
\langle R_\text{ss}^{\prime 2} \rangle + \langle T_\text{ss}^{\prime 2} \rangle = \frac 12 \, ,
\label{12-1}
\end{equation}
where $\langle g(\tau) \rangle = \frac 1{2\pi} \int_0^{2\pi} g(\tau)\dd \tau$, see Appendix \ref{App_A}. 

\subsection{Bilinear spring}
With the same incident wave of \eqref{7}, we have for the interval $\tau\in (\tau_j , \tau_{j+1})$, \ZL{$j\in \mathbb N^+ $,} that $T = f-R $ and  
\beq{8}
R = 
\big [ \sin (\tau +\theta) -  \E^{-(\tau-\tau_j)\tan \theta} \sin (\tau_j +\theta)  \big]
\cos \theta .
\eeq
Note that $\theta$ of eq.\ \eqref{9} depends on the sign of $R$ through eq.\ \eqref{R}.  \ZL{Specifically, 
$\theta =\theta_-$ for $R>0$ and $\theta =\theta_+$ for $R<0$, corresponding to $\alpha=\alpha_-$ and  $\alpha=\alpha_+$, through eq.\ \eqref{9}.}   Since the incident wave is zero for $\tau<0$ it follows that $\tau_1 = 0$  with $\kappa = \kappa_-$ in the interval $\tau\in (\tau_1 , \tau_2)$. 

After several cycles of $f$ the values of $R$ and $T$ become periodic with the same period as $f$.   The value of $T$ varies $2\pi$-periodically between maximum and minimum values defined by two neighboring zero-times of $R$, e.g. $f(\tau_j)$ and $f(\tau_{j+1})$.  The range of the transmission coefficient is therefore contained within the range of $f$, i.e.\ $(-1,1)$. 

\subsection{Transmission design: Unilateral displacement}\label{sec33}

Of interest first are springs with only positive values of transmitted displacement.   In order to simplify the issue we ignore the initial transient and focus on the long-time steady state response which is periodic of period $2\pi$ in terms of the non-dimensional time $\tau$.  Each period comprises a part with $\theta=\theta_-$ and a part with $\theta=\theta_+$.   Let $\tau_\pm$ be the onset time for $\theta_\pm$, then with no loss in generality, taking $\tau_- < \tau_+$, we have $T=f-R$ and 
\beq{9=}
R = \begin{cases}
R_\text{ss}(\tau,\theta_-) + R_\text{tr}(\tau ,\tau_-,\theta_-), 
& \tau \in (\tau_-, \tau_+), 
\\ 
R_\text{ss}(\tau,\theta_+) + R_\text{tr}(\tau ,\tau_+,\theta_+), 
& \tau \in (\tau_+, 2\pi+\tau_-).
\end{cases}
\eeq
By definition, at the onset times the value of $R$ changes sign, and is therefore 
zero, implying 
\beq{34}
\ba
R_\text{ss}(\tau_+,\theta_-) + R_\text{tr}(\tau_+, \tau_-,\theta_-)  &=0 ,
\\
R_\text{ss}(2\pi+\tau_-,\theta_+) + R_\text{tr}(2\pi+\tau_-,\tau_+,\theta_+) &=0  . 
\ea
\eeq

Single sided transmission for which the displacement is non-negative, or equivalently, unilateral transmission, requires as described above that $\tau_- = 0$ \ZL{(by setting this, we can guarantee that the minimum value of $T$ is zero based on eq.\ \eqref{TR})}, and hence the remaining parameters $\tau_+$, $\theta_-$ and $\theta_+$ satisfy
\beq{4-}
\ba
\E^{\tau_+ \tan \theta_-} \sin(\tau_+ +\theta_-) &= \sin \theta_- ,
\\
\E^{-(2\pi- \tau_+ )\tan \theta_+}  \sin(\tau_+ +\theta_+) &=   \sin \theta_+ . 
\ea
\eeq
This describes a one parameter set of bilinear springs which give unilateral transmission.  The set can be parametrized by the turnover time $\tau_+$ when $R$ changes sign   in terms of which  $\theta_+$ and $\theta_-$ are uniquely defined by \eqref{4-}.  The corresponding values of the modified stiffnesses $\alpha_+$ and $\alpha_-$ follow from \eqref{9}. 

Solutions exist for $\tau_+ \in (\frac{\pi}2, \pi)$ with two examples shown in Fig.\ \ref{fig_1}(b)-(c).  The largest range of $T\ge 0$ is obtained for $\tau_+ $ close to but greater than $\frac{\pi}2$ since $T_\text{max} = f(\tau_+)$.  This requires large $\alpha_-$ and small $\alpha_+$, e.g. $\alpha_- /\alpha_+ \approx 23$ for the case in Fig.\ \ref{fig_1}(b). 

Note from \eqref{govern_1} and \eqref{9} that the stiffnesses $\kappa_+$ and $\kappa_-$  scale with the frequency of the incident time harmonic wave.  One can therefore think of  the solutions of \eqref{4-} as defining a unique frequency $\omega$ for a given bilinear spring  such that the transmitted displacement is unilateral
with $T$ strictly positive (negative) if $\kappa_+ < \kappa_-$  $(\kappa_+ > \kappa_-)$.

Given that the transmitted displacement $T$ can be made to be positive or negative, it is natural to ask if the same can be achieved for other quantities.  The reflected displacement $R$ cannot be of a single sign since it is proportional to the  derivative of the periodic function $T$ through eq.\ \eqref{TR}.  However, Fig.\ \ref{fig_1}(b) illustrates that the reflection can be predominantly of one sign, with the values of the opposite sign relatively small.

\section{Transmission through a bilinear spring-mass-spring system } \label{sec4}

\subsection{System assumptions}

\begin{figure}[h]
\centering
\includegraphics[width=0.80\linewidth]{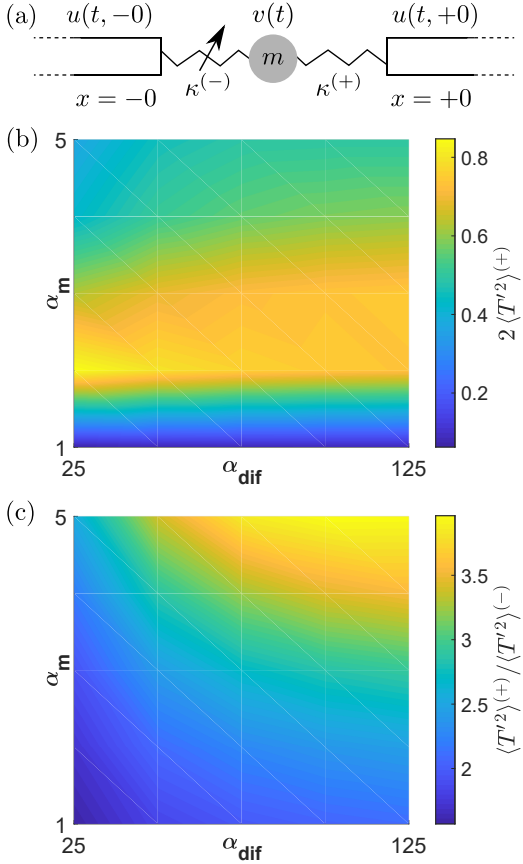}
\caption{Nonreciprocal wave system and numerical results.  (a) Model of two springs with a mass between them.  (b) and (c) Effect of nonreciprocity for different parameter sets in Table \ref{table_asym}.  The \ZL{bilinear stiffness-related} ratio $\alpha_\text{dif} = \alpha^{(-)}_-/\alpha^{(-)}_+$ parameterizes the bilinear spring and \ZL{$\alpha_m$ is a mass dependent parameter (see eqs.\ \eqref{alpha_-} and \eqref{alpha_+m})}; $2 \, \langle T^{\prime \, 2} \rangle ^{(+)}$ is the fraction of energy transmitted, and $\langle T^{\prime \, 2} \rangle ^{(+)} / \langle T^{\prime \, 2} \rangle ^{(-)}$ is a measure of nonreciprocal energy transmission.}
\label{fig_2}
\end{figure}

The significant diode-like transmission happens when we introduce spatial asymmetry to the gap between $x=-0$ and $x=+0$.  This can be simply realized by adding another linear spring and an additional mass to the previous case.  From now on, the coupling between the waveguides is a simple spring-mass chain system of mass $m$, a bilinear spring $\kappa^{(-)}$ and a linear spring $\kappa^{(+)}$, see Fig.\ \ref{fig_2}(a).   

\begin{table}[ht]
\centering
 \begin{tabular}{|| c | c | c | c | c ||} 
 \hline
 \ZL{$\omega$} & $\alpha^{(-)}_-$ & $\alpha^{(-)}_+$ & $\alpha^{(+)}$ & $\alpha_m$  \\
 \hline\hline
 \ZL{$1$} & $5 \sim 25$ & 0.2 & 0.5 & $1 \sim 5$ \\
 \hline
\end{tabular}
\caption{Wave frequency (unit of $rad/s$) and the parameters \ZL{(units of $1/s$ see eqs.\ \eqref{alpha_-} and \eqref{alpha_+m})} considered for the asymmetric nonreciprocal system of Fig.\ \ref{fig_2}(a).  }
\label{table_asym}
\end{table}

Let $\w = \w(t)$ be the mass displacement, so that the equilibrium equation for the mass is 
\beq{u0}
m \w'' = F(t, +0) - F(t, -0) \, ,
\eeq
where the spring forces are 
\beq{F_2}
F(t, \pm 0) = \kappa^{(\pm)} \, [ u^{(\pm)} ] \, ,
\eeq
with  extensions (see Eq.\ \eqref{u_xpm}) 
\beq{u2_pm}
\ba
{[u^{(+)}]} & = T - \w \, , 
\\
[u^{(-)}] & = \w - R - f \,  . 
\ea
\eeq
Alternatively, using Eqs.\ \eqref{u_xpm} and \eqref{F2} allows us to write 
 the mass equilibrium equation  as 
\beq{u0_2}
m\w''(t) =
-\frac{EA}c \big(T'+R'-f' \big) \, ,
\eeq
which may be integrated once.  Combined with the two equations for $F^{(\pm)}$ in terms of $R$, $T$ and $f$, we obtain a system of three ODEs for the unknowns $R$, $T$ and $\w$ in terms of the incident wave $f$:
\bse{8-5}
\bal{8-5a}
R' &= \frac 12 \alpha^{(-)} \big( \w-R-f \big) + f', 
\\
T' &= \frac 12 \alpha^{(+)} \big( \w-T \big) , \label{8-5b}
\\
\w' &= \alpha_m \big(f-R-T \big), 
\eal
\ese
where following Eqs.\ \eqref{govern_1} and \eqref{alpha}, since there are now a bilinear spring and a linear one, we have  
\beq{alpha_-}
\alpha^{(-)} = \frac{2c \kappa^{(-)}}{EA} = \begin{cases}
\alpha_-^{(\pm)}, & [u^{(-)}]<0 , 
\\
\alpha_+^{(\pm)}, & [u^{(-)}] >0,
\end{cases}
\eeq
and 
\beq{alpha_+m}
\alpha^{(+)} = \frac{2c \kappa^{(+)}}{EA} \, , \, \alpha_m  \equiv \frac{EA}{mc} .
\eeq
The corresponding equations for incidence from the right are the same as \eqref{8-5} but with $\alpha^{(-)}$ and $\alpha^{(+)}$  swapped between \eqref{8-5a} and \eqref{8-5b}. 

\subsection{Nonreciprocal transmission}

We  quantify wave nonreciprocity from the perspective of energy flux by finding the transmission coefficients for incidence from opposite directions: $2 \, \langle T^{\prime \, 2} \rangle ^{(\pm)}$, where $^{(-)}$ means incidence  from the left and $^{(+)}$ from the right.   A large transmission coefficient indicates  high transmission capability, and the ratio of two transmission coefficients is a measure of nonreciprocal wave propagation.  Here we assume that the transmission coefficient for propagation from the right is the greater one of the two in terms of energy transmission: $\langle T^{\prime \, 2} \rangle ^{(+)} >  \langle T^{\prime \, 2} \rangle ^{(-)}$,  Therefore,  significant nonreciprocal transmission occurs when both $2 \, \langle T^{\prime \, 2} \rangle ^{(+)}$ and $\langle T^{\prime \, 2} \rangle ^{(+)} / \langle T^{\prime \, 2} \rangle ^{(-)}$ are as large as possible. 

Extreme bilinearity and asymmetry can be realized with one stiffness of the bilinear spring much larger than the other stiffness.  Here we assume the stiffness of the bilinear spring is much greater in compression than  in tension and also much greater than  the linear spring stiffness: $\alpha^{(-)}_- \gg \alpha^{(+)} > \alpha^{(-)}_+$.  The mass in the middle also has a strong effect on the nonreciprocity.  Based on these observations we consider  the parameters in Table \ref{table_asym}, with associated numerical results in Fig.\  \ref{fig_2}(b) and \ref{fig_2}(c).


\subsection{Dynamic analysis}

\begin{figure}[ht!] 
\centering
\includegraphics[width=0.8\linewidth]{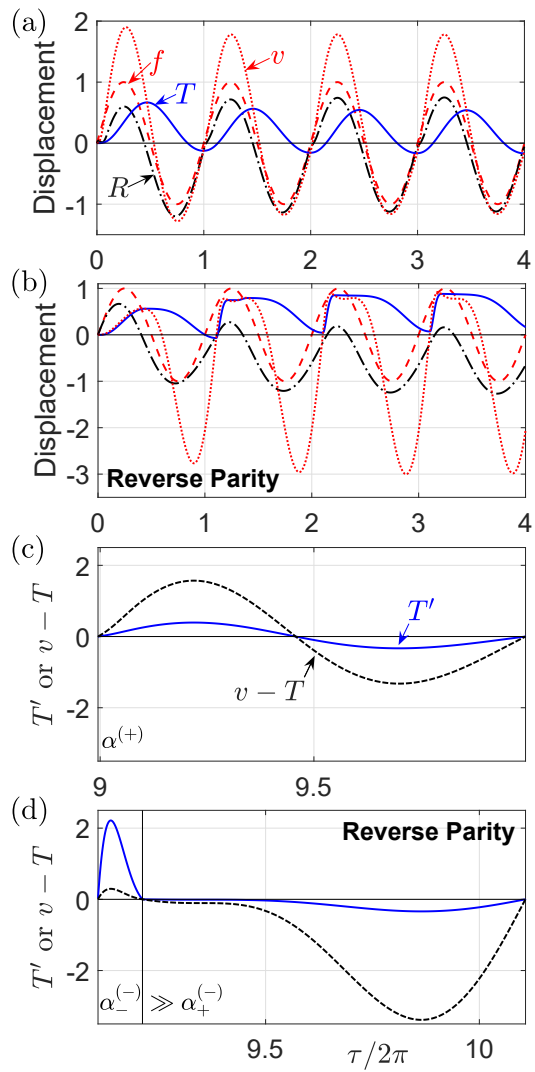}
\caption{Nonreciprocal dynamic properties for the model of Fig.\ \ref{fig_2}(a) with parameters $(\alpha^{(-)}_-, \alpha^{(-)}_+, \alpha^{(+)}, \alpha_m) =$ $ (15, 0.2, 0.5, 5)$. For (a) and (c) the incident wave $f$ (red dashed line) is from the left, and for  (b) and (d) from the right. (a) and (b) show the reflected wave $R$ (black dashdotted lines), transmitted wave $T$ (blue line) and mass displacement $\w$ (red dotted line) vs.\ time.  \ZL{(c) and (d) depict the changes of $T^{\prime}$ and $v - T$ over one period; Also, the corresponding $\alpha^{(\pm)}$ values are labeled (the intervals are separated by the vertical black lines in (d) according to the bilinear stiffness property).  The significant nonreciprocity is thought to be realized in this case because of (1) the obvious difference in the dynamic behaviors of (a) and (b), and (2) the small time interval with large values of the energy-related parameter $T^{\prime \, (+)}$ occurring only in (d) instead of (c).}  }
\label{fig_behavior}
\end{figure}

Figure \ref{fig_behavior} presents results for a particular model that displays a huge difference in transmission properties for incidence from opposite directions. 
In this case, we have $2 \, \langle T^{\prime \, 2} \rangle ^{(+)} \approx 50 \%$ and $\langle T^{\prime \, 2} \rangle ^{(+)} / \langle T^{\prime \, 2} \rangle ^{(-)} = 3.85$.  Figure \ref{fig_behavior}(a) and (c) show the dynamic properties for  incidence  from the left, and Fig.\ \ref{fig_behavior}(b) and (d) from the right. 
It is clear that the transmitted wave $T$, reflected wave $R$ and the displacement of the central mass $\w$ have different responses for the different incident directions. 
The behaviors of $T^{\prime}$ vs.\ time, which are relevant to the transmitted energy flux, also show a significant difference for incidence from opposite directions.  Let us examine these time histories more closely.

From the perspective of displacement, when incidence is from the left, the spring on the left is compressed first. Since the stiffness in compression is the largest among all stiffnesses ($\alpha^{(-)}_- \gg \alpha^{(-)}_+ , \alpha^{(+)}$), the mass in the middle is pushed to the right direction with large  displacement as Fig.\ \ref{fig_behavior}(a) shows. The spring on the right is then compressed, which causes  the transmitted wave to have positive displacement. Once the incident forcing puts the left spring in tension, the mass is drawn back and moves to the left  but with  smaller displacement. The transmitted displacement has  the same behavior.
Similarly, when incidence is from the right, the spring on the right is compressed first. The mass in the middle is pushed to the left  and the spring on the left is compressed. However, the mass displacement to the left  is small because the compressive stiffness of the spring on the right is small and that of the spring on the left is large. Then the mass has a drastic move towards the right with larger displacement because of the the small tensile stiffness of the left and right springs.

From the perspective of energy flux, the transmitted energy flux $\langle T^{\prime \, 2} \rangle$ mainly depends on the values of $\alpha^{(+)}$ for incidence from the left and $\alpha^{(-)}$ from the right, as the second expression in Eq.\ \eqref{8-5} shows.  Since $\alpha^{(-)}_- \gg \alpha^{(-)}_+$, the significant difference between the $T^{\prime}$ vs.\  time for incidence from the opposite directions is clear in  Figs.\ \ref{fig_behavior}(c) and (d): the relatively small time interval with large values of  $T^{\prime \, (+)}$ in Fig.\ \ref{fig_behavior}(d) results in the large energy flux $\langle T^{\prime \, 2} \rangle ^{(+)}$; However, this phenomenon does not  occur in Fig.\ \ref{fig_behavior}(c); and we therefore  get the nonreciprocal energy flow $\langle T^{\prime \, 2} \rangle ^{(+)} > \langle T^{\prime \, 2} \rangle ^{(-)}$.

\subsection{Semi-analytical solution}

The dynamics of a bilinear spring can be described by piecewise linear solutions patched together at the instants the bilinear stiffness changes. Using this observation we show that the  steady state dynamic behavior of the system in Fig.\ \ref{fig_2}(a) can be solved using semi-analytical methods. 

Since the time harmonic incident wave $f = \sin \omega  t$, \ZL{the system Eq.\ \eqref{8-5} can be simplified in  matrix form
\begin{equation}
\boldsymbol{V} (t) ' = \boldsymbol{M} \, \boldsymbol{V} (t) + \Re \big [ \boldsymbol{F} \, e^{-i \omega  t} \big ] \, ,
\label{v_mat}
\end{equation}
where time dependent and constant vectors are
\begin{equation}
\boldsymbol{V} (t)= 
\begin{pmatrix}
R \\
T \\
\w
\end{pmatrix} \, , \,
\boldsymbol{F} = \frac{1}{2} \,
\begin{pmatrix}
2 \, \omega - i \, \alpha^{(-)} \\
0 \\
i \, 2 \, \alpha_m  
\end{pmatrix} \, ,
\end{equation}
with  constant matrix
\begin{equation}
\boldsymbol{M} = \frac{1}{2} \, 
\begin{pmatrix}
- \alpha^{(-)} & 0 & \alpha^{(-)} \\
0 & - \alpha^{(+)} & \alpha^{(+)} \\
- 2 \, \alpha_m & - 2 \, \alpha_m & 0
\end{pmatrix} \, .
\end{equation}
}The solution of the system Eq.\ \eqref{v_mat} within a given piecewise linear interval can be written as  the sum of a homogeneous solution  $\boldsymbol{V}^{(1)} (t)$ plus a particular solution: 
\begin{equation}
\boldsymbol{V} = \Re \big [ \boldsymbol{V}_{(0)} \, e^{-i \omega  t} \, \big ] + \boldsymbol{V}^{(1)} \, ,
\label{v}
\end{equation}
where \ZL{$\boldsymbol{V}_{(0)}$ is the constant coefficient vector.  Note that $\boldsymbol{V}$ with the subscript denotes a constant vector and with the superscript a time-dependent vector.}

The vector $\boldsymbol{V}_{(0)}$ associated with the particular solution satisfies
\begin{equation}
( i \omega \, \boldsymbol{I} + \boldsymbol{M} ) \, \boldsymbol{V}_{(0)}
=
- \boldsymbol{F} \, .
\end{equation}
Since $(i \omega \, \boldsymbol{I} + \boldsymbol{M})$ is invertible, it follows that 
\begin{equation}
\boldsymbol{V}_{(0)} = - \big ( i \omega \, \boldsymbol{I} + \boldsymbol{M} \big ) ^{-1} \, \boldsymbol{F} \, . 
\label{v0}
\end{equation}

\begin{figure}[ht!] 
\centering
\includegraphics[width=0.8\linewidth]{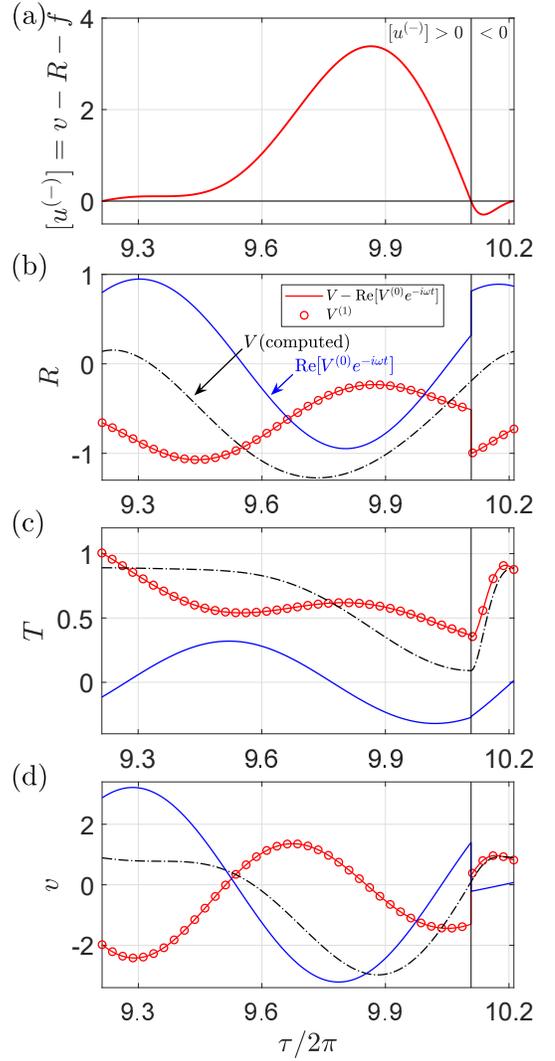}
\caption{Comparison of numerical and analytical results.  (a) shows the sign of $[u^{(-)}]$, indicating that the dynamic process over one period is split into two intervals of piecewise constant stiffness. Intervals are separated by the vertical black lines in (b) - (d) which show the reflected wave $R$, transmitted wave $T$ and displacement of central mass $\w$, respectively. The black dashdotted lines depict the computed \ZL{numerical results} $\boldsymbol{V}$; the blue lines are the pure analytical solution (the \ZL{particular solution} in eq.\ \eqref{v}), $\Re [ \boldsymbol{V}^{(0)} \, e^{-i \omega t} ]$; the red solid lines show the remainder $\boldsymbol{V} - \Re [ \boldsymbol{V}^{(0)} \, e^{-i \omega t} ]$; the red dots show the semi-analytical results of this remaining part (the \ZL{homogeneous solution} part in eq.\ \eqref{v}) $\boldsymbol{V}^{(1)}$.  The calculated and semi-analytical results match well over the full period, verifying eq.\ \eqref{V_eqn}. }
\label{interval_strong_1}
\end{figure} 

The homogeneous solution in Eq. \eqref{v} satisfies
\begin{equation}
\boldsymbol{V}^{(1) \, \prime} = 
\boldsymbol{M} \, \boldsymbol{V}^{(1)} \, ,
\end{equation}
and the solution can be expressed as
\begin{equation}
\boldsymbol{V}^{(1)} = \boldsymbol{V}_{(1)} + e^{\boldsymbol{M} \, (t - t_j)} \, \boldsymbol{V}_{(2)} \, ,
\label{V_1t}
\end{equation}
where both $\boldsymbol{V}_{(1)}$ and $\boldsymbol{V}_{(2)}$ are constant, and $t_j$ is the starting time of interval $j$. The first constant vector satisfies $\boldsymbol{M} \, \boldsymbol{V}_{(1)} = \boldsymbol{0}$, implying $\boldsymbol{V}_{(1)} = \boldsymbol{0}$ and 
\begin{equation}
\boldsymbol{V}^{(1)} = e^{\boldsymbol{M} \, (t - t_j)} \, \boldsymbol{V}_{(2)} \, .
\label{v1_1}
\end{equation}
The remaining vector can be found by the initial condition for each interval at the start time  $t = t_j$, \ZL{such that
\begin{equation}
\boldsymbol{V}_{(2)} = \boldsymbol{V}^{(1)} (t_j) \, ,
\end{equation}
the exact value of which can only be found numerically from }
\begin{equation}
\boldsymbol{V}^{(1)}(t_j) = \boldsymbol{V} (t_j) - \Re \big [ \boldsymbol{V}_{(0)} \, e^{-i \omega  t_j} \, \big ] \, .
\label{v1_2_value}
\end{equation}

In summary,  Eqs.\ \eqref{v}, \eqref{v0}  and \eqref{v1_1}-\eqref{v1_2_value} imply 
\begin{align}
&\boldsymbol{V} = 
- \Re \big [ \big ( i \omega \, \boldsymbol{I} + \boldsymbol{M} \big ) ^{-1} \, \boldsymbol{F} \, e^{-i \omega  t} \, \big ] + 
  \notag\\
& e^{\boldsymbol{M} \, (t - t_j)} \Big ( \boldsymbol{V} (t_j) + \Re \big [ \big ( i \omega \, \boldsymbol{I} + \boldsymbol{M} \big ) ^{-1} \, \boldsymbol{F} \, e^{-i \omega  t_j} \, \big ] \Big ) \, .
\label{V_eqn}
\end{align}

Figure \ref{interval_strong_1} shows an example of applying the semi-analytical method to the case shown in Fig.\ \ref{fig_behavior}(b) for incidence from the right.  Figure \ref{interval_strong_1}(a) shows the sign of $[u^{(-)}]$ vs.\ time indicating that the dynamic process over one period comprises two linear intervals in which the \ZL{stiffness related parameter $\alpha^{(-)}$} is piecewise constant.  \ZL{Only one period is plotted starting with $[u^{(-)}] = 0$, the first interval of which is $[u^{(-)}] > 0$ and the second $[u^{(-)}] < 0$.}  The intervals are separated by the vertical black lines.  Figures \ref{interval_strong_1}(b) to (d) depict the comparison of \ZL{the semi-analytical homogeneous solution $\boldsymbol{V}^{(1)}$ and the difference between the computed numerical results and pure analytical particular solution $\boldsymbol{V} - \Re [ \boldsymbol{V}_{(0)} \, e^{-i \omega  t} ]$.}  These two sets of values match very well in both intervals. Figures \ref{interval_strong_1}(b), (c) and (d) depict information for the reflected wave $R$, transmitted wave $T$ and displacement of central mass $\w$, respectively.

\section{Conclusion} \label{sec5}

Using a single bilinear spring element in an otherwise linear system we have demonstrated the possibility of passive amplitude independent nonreciprocal wave effects.  The amplitude independence means that the output signal scales linearly with the input.  While the bilinear spring provides the necessary nonlinear property for achieving nonreciprocity, the sufficient condition of spatial asymmetry is obtained using a single linear spring to offset the bilinear one.  Significant nonreciprocity is observed when the bilinearity is strong in the sense that the stiffnesses in compression and in tension are highly dissimilar.  The nonreciprocal wave system is amenable to a semi-analytic solution that takes advantage of the piecewise linear nature of the dynamics. This property also makes the system unique among passive nonreciprocal wave systems.  When the connection between the waveguides is a single bilinear spring the system becomes reciprocal, although it can still display interesting wave effects, such as the  phenomenon of unilateral transmission discussed here for the first time.  

\subsubsection*{Acknowledgment} This work is supported by the NSF EFRI Program under Award No. 1641078. 
 
\bigskip
\appendix

\section{Energy considerations} \label{App_A}
The dynamic equation for $u(t,x)$,  
\beq{04}
E Au_{,xx} - \rho A  u_{,tt} = 0
\eeq
becomes, after multiplication by velocity $u_{,t}$,
\beq{041}
\partial_t {\cal E}  +  \partial_x {\cal F}  = 0
\eeq
where ${\cal E}$ and ${\cal F}$ are the energy density and energy flux
\beq{043}
{\cal E} =\frac 12 A\big( Eu_{,x}^2 + \rho u_{,t}^2 \big), 
\ \
{\cal F} = - EA u_{,x} u_{,t}.
\eeq
For a traveling wave, e.g. $u = T(t-x/c)$ the flux reduces to 
\beq{044} 
{\cal F} = \rho c A {T'}^2.
\eeq
In the steady state limit the associated energy is the average of the flux over one period.


\end{document}